# Giant solitary fibrous tumor of the pleura

*Javier Alfonso Pinedo-Onofre,\* Eurídice Robles-Pérez,\* Erika Sagrario Peña-Mirabal,\*\* José Amado Hernández-Carrillo,\* and José Luis Téllez-Becerra\**

**Abstract**

**Background**: Solitary fibrous tumor is the second primary malignancy of the pleura and can reach up to 39 cm in diameter; however, to be referred to as "giant" it must occupy at least 40% of the affected hemithorax. Although this tumor usually shows a benign behavior, malignancy criteria have been described. The aim of the study was to assess the initial evaluation, diagnostic procedures, surgical management, treatment outcome, and prognosis.
**Methods**: We performed a descriptive, observational, longitudinal, and retrospective study from 2002 to 2006 on patients who underwent surgery with a diagnosis of giant solitary fibrous tumor of the pleura.
**Results**: Six patients were included; 83.3% were females. Mean age was 48 years. All patients were symptomatic, mainly dyspnea, cough and chest pain; 66.7% were left-sided. Preoperative angiography and embolization were performed in 83.3% cases with successful surgical resection. The predominant blood supply was derived from the internal mammalian artery. Intraoperative complication rate was 17%. A vascular pedicle was found in 66.7%. The largest lesion was 40 cm in diameter and weighed 4500 g. Only one case showed high mitotic activity. Mean follow-up to date is 14 months.
**Conclusions**: Symptomatology found was consistent with previous reports but in higher percentages. Accurate diagnosis is critical because surgical resection involves a potential cure; however, long-term follow-up is mandatory. Preoperative embolization is recommended due to tumor size.

## Introduction

Most pleural neoplasms have a metastatic origin. Primary pleural tumor can be categorized as diffuse (diffuse malignant mesothelioma) or localized (previously referred to as localized mesothelioma and now referred to as "fibrous solitary tumor" or "localized fibrous tumor of the pleura"); the latter has a controversial histogenesis.[1-3]

Localized fibrous tumor of the pleura was first described by Wagner in 1870; however, it was not until 1931 when Klemperer and Rabin described it pathologically as a separate entity.[3-5] This entity has received diverse names such as localized mesothelioma, benign fibrous mesothelioma, fibrous solitary mesothelioma, localized benign fibroma, submesothelial fibroma, pleural fibroma and subserous fibroma.[6-8]

Approximately 800 cases have been reported in the international literature, and this entity has frequently been confused with pleural mesothelioma.[3,9] This disease represents <5% of pleural neoplasms with a prevalence of ~2.8/100,000 tumors.[4,10-12] Some authors report that it constitutes 8% of benign thoracic neoplasms and 10% of pleural tumors,[6] being the second most important pleural primary tumor after malignant mesothelioma[10] and representing <1% of the total surgical activity of the thorax.[4]

This disease affects both males and females without racial specificity and is more frequent during the sixth and seventh decades (although it may affect any age group).[1-3,7,13,14] Some authors have reported a higher affection in males[15,16] as well as a family relationship between mother and daughter.[3] Other authors have reported an associated involvement from chromosomes 8 and 12,[17-20] particularly trisomy 8 and trisomy 21, finding a relationship between tumor size and genetic abnormalities. This suggests that these chromosomal defects can promote tumor growth.[21]

\* Subdirección de Cirugía,
\*\* Servicio de Anatomía Patológica, Instituto Nacional de Enfermedades Respiratorias, México, D.F., Mexico

*Correspondence and reprint requests to*
Javier Alfonso Pinedo-Onofre
Instituto Nacional de Enfermedades Respiratorias
Subdirección de Cirugía
Calz. de Tlalpan 4502, Col. Sección XVI, Del. Tlalpan
14080 México, D.F.
Tel.: (55) 5666 8110; Fax: (55) 5666 0997
E-mail: dr_creatura@hotmail.com

Fibrous tumors of the pleura are solitary, although extremely rare cases with synchronous lesions have been reported[22] as well as intrapulmonary lesions[17] (7.5% of cases[23,24]). The latter are referred to as inverted fibromas[1] and are associated with these two characteristics.[6] The tumor is generally encapsulated, measuring on average 5-6 cm in diameter with reports including 1-39 cm in diameter[1] and with a weight up to 5200 g.[25] The tumor may also be lobulated.[26]

"Giant" tumors receive this name when they occupy at least 40% of the affected hemithorax.[27] They have a fibrous grayish-white surface[1,13,26] alternating with soft areas, necrosis and hemorrhage. Sections reveal nodules composed of densely agglutinated fibrous tissue and may contain cystic structures from several millimeters to 5 cm in diameter with clear viscous liquid in up to 10-15% of cases.[1,13] A short peduncle has been described in 38-50%,[2,13,24] and is about <1 cm in length.[13] In up to 66% of cases, this peduncle attaches to the visceral pleura[13,24,28] and can be vascularized in larger tumors, associated with increased vascularity.[1,8] The presence of a peduncle usually indicates a benign tumor, independent of histological characteristics[7] that show formation from fusiform homogeneous cells in a densely hyalinized fibrous stroma and rarely calcified (0.9%).[13] Atypical cell formations and mitosis are infrequent.[2,29]

These tumors originate due to mesenchymatous growth produced from visceral or parietal pleura.[4,9] Tumors from visceral origin are more frequent,[2] representing >80% of cases.[11] Immunohistochemical tests have confirmed that these tumors do not have a mesothelial origin[4,7,28] but are more likely to originate from primitive mesenchymatous submesothelial cells.[2,3,28,30] These lesions show a low prevalence on the right hemithorax.[31] Tumors developing in atypical places are malignant (thoracic wall parietal pleura, diaphragm, mediastinum, lung fissures, intraparenchymatous).[1,13] Because solitary fibrous tumors can originate in peritoneum, pericardium, mediastinum, meninges, lung, heart, thyroid, parotid glands, kidney, adrenal glands, bladder, ocular orbit, oral cavity, epiglottis, nose and paranasal sinuses, it has been suggested that these entities may originate from a common stem cell present in several organs and tissues and possibly have a myofibroblastic origin.[5-8,11,32-34]

## Patients and Methods

We carried out a descriptive, observational, longitudinal and retrospective study between January 1, 2002 and December 31, 2006, which included all patients intervened and diagnosed with giant fibrous solitary tumor of the pleura at the National Institute of Respiratory Diseases "Dr. Ismael Cosio Villegas" in Mexico City. The purpose of our study was to report the experience obtained in management of this rare pathology as well as to carry out an extended review of diagnostic and therapeutic approaches based on a selective literature review. Descriptive statistical analysis was carried out using basic statistics.

## Results

Of seven cases found, we excluded one because of the final diagnosis as monophasic synovial sarcoma. Of six cases included in our study, 83.3% ($n = 5$) were females with an average age of 48 years (range: 30-72 years) without prevalence in any age group. Exposure to asbestos or tobacco was not documented in any case. Only systemic arterial hypertension was observed in one case (16.7%) as a comorbidity. All patients were symptomatic, presenting dyspnea, cough and thoracic pain as the most frequent symptoms (Figure 1).

Posteroanterior chest x-ray was carried out in all cases, estimating tumor volume at an average 70% of affected hemithorax (range: 50-95%), observing contralateral mediastinal deviation in all cases dependent on tumor size. Computed tomography (CT) of the chest was carried out in all patients without clearly identifying vascular peduncle or invasion to other structures. Magnetic resonance imaging (MRI) studies were not carried out.

Fine needle aspiration biopsy (FNAB) was carried out in only one case; however, this was not diagnostically useful. In three cases (50%) true-cut biopsy was carried out, resulting in a diagnosis in 66.7% of cases ($n = 2$).

In 66.7% ($n = 4$) of cases, the tumor was located on the left side. In four cases the tumor also had a parietal pleura origin showing a prevalence on the mediastinal pleura in three cases, whereas the other two cases were diaphragmatic with only 33.3% originating from visceral pleura.

Preoperative angiography and embolization were carried out in 83.3% ($n = 5$) of cases, all of them female, identifying arterial contribution with a prevalence of internal mammary artery (100%, $n = 5$) as well as bronchial arteries (60%, $n = 3$) and phrenic arteries (40%, $n = 2$).

Embolization was carried out with polyvinyl alcohol microspheres in all cases. Only one case reported paresis of lower right limb from medullary ischemia, which receded 2 weeks later without further consequence, resulting in a 20% morbidity rate attributable to procedure. All embolized cases underwent surgical resection at an average of 7 days after procedure (range: 2-18 days). The range reaches 18 days because of medullary ischemia presented in one case.

Complete tumor resection was carried out in all embolized cases ($n = 5$) through posterolateral thoracotomy approach. The other case was partially resected (demon-strated through microscopy). In case of complete tumor resection, an additional pulmonary parenchyma V-shaped resection was carried out. Average intraoperative bleeding was 1908.3 ml (range: 650-3300 ml). The patient who had the highest amount of bleeding (3300 ml) required packing of the thoracic cavity that was resolved in 24 h, resulting in an intraoperative complication rate of 16.7%. There was zero mortality. No case required adjuvant therapy.

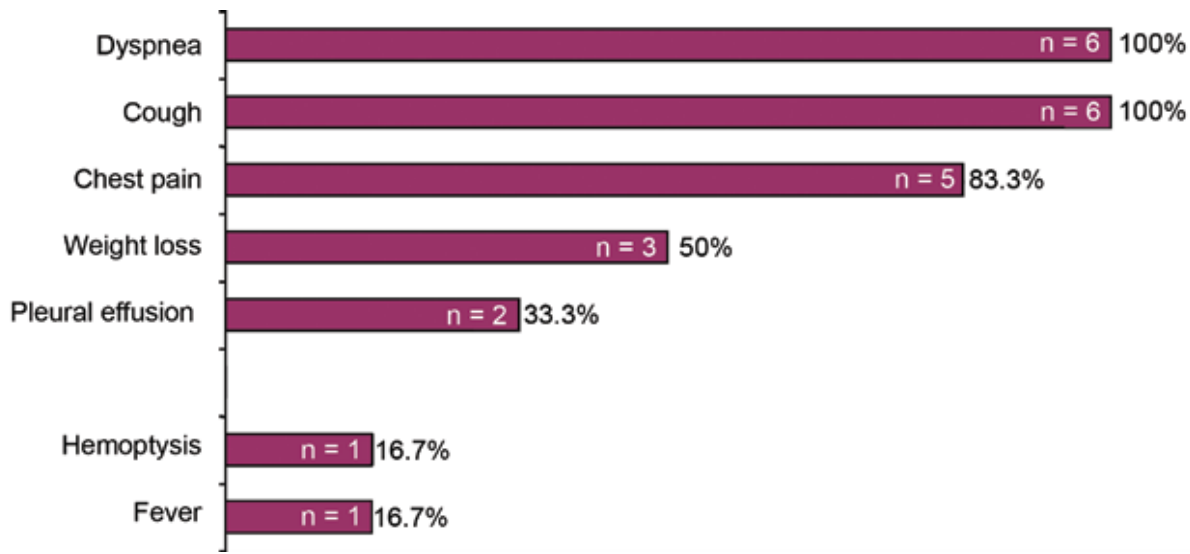

**Figure 1.** Symptoms reported in studied cases.

All tumors were solitary with an average size of 29.3 cm (range: 24-40 cm) at their largest diameter with an average weight of 2567 g (range: 1700-4500 g). In 66.7% (*n* = 4) of cases, a vascular peduncle was identified, three were dependent on the parietal pleura and the other on the visceral pleura with an average diameter of 1.75 cm (range: 1-3 cm). All lesions were encapsulated and none presented evidence of invasion to other structures. High mitotic activity was identified in only one case (16.7%) (>4 mitosis per 10 high-power fields) that corresponded to a malignant tumor. Table 1 presents macro- and microscopic findings and Figure 2 illustrates preoperative tests and postoperative x-rays of case 3.

Immunohistochemical tests were negative in all cases for cytokeratins, bcl-2, SMA, EMA, S-100 protein, desmin and calretinin and positive in all cases for CD34 and for vimentin in 83.3% of cases.

Average follow up was 14 months (range: 2-33 months). One relapse was documented in the nonembolized patient who also experienced incomplete tumor resection. Relapse occurred 4 months after follow-up, demonstrating a 16.7% relapse rate. The patient was reintervened due to a tumor in the apex of the lung and fourth costal arch. Histopathological tests reported high-grade pleomorphic sarcoma compatible with giant-cell malignant fibrous histiocytoma, resulting in a malignant transformation rate of 16.7%.

## Discussion

Solitary fibrous tumors of the pleura may be benign or malignant when they diffusely infiltrate lung parenchyma[29] or adjacent structures.[1] The benign variant is an encapsulated lobulated tumor without a microscopically defined pattern and may appear as a sarcomatoid mesothelioma with inserted fascicles from ovoid or fusiform cells with poorly defined borders and scarce cytoplasm, without atypia, and with cells inversely proportional to collagen.[7,13] Malignant variant (~10-33% of cases) is characterized by high cellularity, accented pleomorphism (expressed as increased nuclear grade) and high mitotic activity (>4 mitosis per 10 high-power fields[7,13]),[1,4,9,11,35-37] and usually presents necrosis or hemorrhage.[1-2] These five histological criteria plus the possibility of pleural effusion, atypical location and adjacent structure invasion have been established as malignant indicators by England et al.[13] (Table 2) and accepted by the American Registry of Pathology;[38-40] however, only mitotic index has shown a statistically significant association with poor biological behavior,[6] as we observed in our study.

Only one of England's criterion is required, even focused, to regard the tumor as malignant.[24] We did not consider the presence of necrosis and hemorrhage to classify malignancy because of the time elapsed between preoperative embolization and surgical procedure (average: 7 days; range: 2-18 days) because the literature mentions that this period should be reduced to hours. Even though most cases are reported as benign, malignant behavior has been reported in long-term

**Table 1.** Macro- and microscopic findings in analyzed cases

|  | Case 1 | Case 2 | Case 3 | Case 4 | Case 5 | Case 6 |
| --- | --- | --- | --- | --- | --- | --- |
| Age | 30 | 46 | 50 | 54 | 38 | 72 |
| Gender | F | F | F | F | F | M |
| Embolization | Yes | Yes | Yes | Yes | Yes | No |
| Full resection | Yes | Yes | Yes | Yes | Yes | No |
| Size (cm) | 27 x 20 x 7 | 40 x 30 x 15 | 24 x 20 x 9 | 24 x 22 x 22 | 32 x 26 x 18 | 29 x 17 x 9 |
| Weight (g) | 2500 | 2000 | 1700 | 2700 | 4500 | 2000 |
| Invasive | – | – | – | – | – | – |
| Encapsulated | + | + | + | + | + | + |
| Necrosis | + | + | – | – | + | – |
| Hemorrhage | + | + | – | – | + | – |
| Cystic contents | – | – | – | Mucous | – | – |
| Histological pattern | Mixed | Hyaline degeneration | Fibrous hyaline | Mixed | Mixed | Mixed |
| Calcification | – | – | – | + | – | – |
| Atypia | Mild | – | Mild | Mild | Mild | – |
| Cellular density | Intense | Intens | Moderate | Moderate | Moderate | Moderate |
| Mitosis | 0 | 0 | 1-2 | 1-4 | 1-4 | 1-5 |
| Free borders | Yes | Yes | Yes | Yes | Yes | No |
| Classification | Benign | Benign | Benign | Benign | Benign | Malignant |

M, male; F, female; +, Present; –, Absent.

follow-up,[7,37] which is not always referred to histological criteria of malignancy[41] (Table 2). The rate of benign vs. malignant tumors has been reported as 7:1[15,40] with an incidence of invasive behavior in 13-23% of cases.[32]

Contrary to mesotheliomas, solitary fibrous tumors of the pleura report negative immunohistochemical results for cytokeratins[3,9] (although malignant variations occasionally report positive results[41] and express through submesothelial cells under certain circumstances[31]), positive to vimentin and CD34[3,4,24] (Table 3), positive in 29-50% of cases to Kit (CD117),[35,42] and have no relationship with exposure to asbestos[2,4,35] or tobacco.[2] However, they can present an unpredictable biological behavior, sometimes with local recurrence and metastases even with negative CD34. These negative results may be useful for biological prognosis.[6,14] It has been reported that a strong p53 expression with negative CD34 and high Ki67 index (proliferative index) are associated with poor biological behavior (Table 2).[36,43,44]

The Ki67 antigen is associated with cell proliferation and is part of DNA replicase complex. Its presence indicates proliferative activity in diverse tumors and in contrast with proliferative cellular nuclear antigen (PCNA), which intervenes also in DNA repair. It is virtually restricted as a proliferation antigen, which provides a better indicator of growth fraction. MIB1 is a monoclonal antibody developed against Ki67 antigen.[45] Proliferative index (Ki67/MIB1) indicates the number of dividing (proliferating) cells in a tumor and can be used jointly with S-phase fraction to provide a better understanding of the rate of tumor growth.[21,46] It is expressed as the percentage of stained nuclei in a sample[47] (0-10% in benign tumors, 20-40% in malignant tumors[21]). There are contradictory reports about the expression of estrogen and progesterone receptors and their association with malignancy because of the small sample size; therefore, further investigation is required.[41,48] It has been suggested that tumor growth may be accentuated during the last trimester of pregnancy possibly because of increased hormonal activity.[49]

From revised series, only England reports positive results for vimentin in 81% of cases, whereas the others report 100% positive results (Tables 3 and 4). England's series included 223 patients and we found positive results in 83.3% of cases according to England's findings despite the much smaller sample size.

Vimentin, desmin and keratins are cytoskeletal intermediate filaments (proteins). Vimentin is expressed in all mesenchymatous cells that cover (like an epithelium) serous membranes and in fibroblasts, fibrocytes and vascular endothelial cells. Desmin is associated with skeletal and smooth muscle differentiation and keratin is a marker for epithelial differentiation.[50,51] S-100 protein is a marker of peripheral nerve tumors.[50] CD99 is produced by the MIC2 gene and intervenes in regulation of cellular proliferation and adhesion.[50] Calretinin has a high expression in neurons and external nervous system in mesothelial cells; therefore, it is used as a mesothelioma marker in differential diagnosis involving carcinoma.[52]

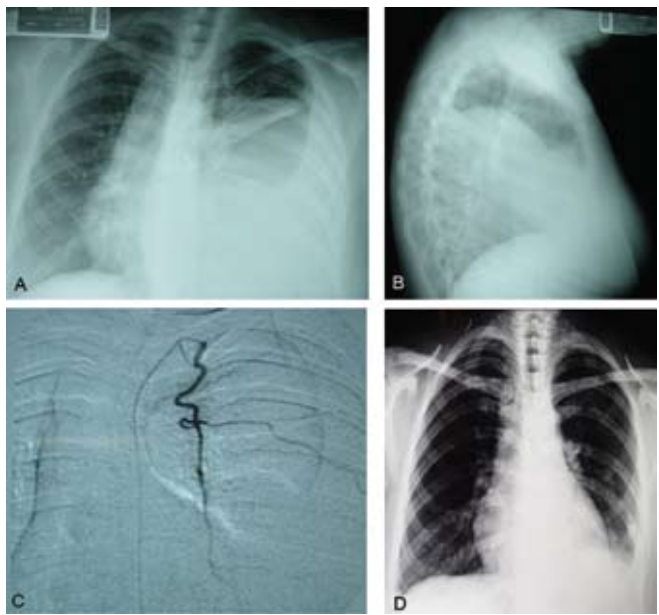

**Figure 2.** (A,B) Posteroanterior and lateral chest x-rays showing homogeneous opacity in left hemithorax with well-defined borders that occupies about 60% of hemithorax. Phrenic sinuses are blurred and contralateral mediastinum deviation. (C) Angiography showing tumor irrigation through left internal mammary artery. (D) Control x-ray 2 weeks after surgery.

CD34 is a transmembrane protein present in human hematopoietic progenitor cells and in vascular endothelium. It acts as an ubiquitous antigen in a wide range of histogenic cellular types without having a specific lineage and is expressed by myofibroblasts. Therefore, it is not specific for these tumor types and should be interpreted in an appropriate histomorphological context in combination with other markers.[7,14] Because of its myofibroblast origin, smooth muscle actin (SMA) staining is generally negative because there are four myofibroblast types and one type does not express actin.[7] Kit (CD117) is a transmembrane receptor with tyrosine kinase activity produced by proto-oncogene *c-kit*, which regulates development and growth of certain human cell types and has been found in pulmonary and pleural tumors. It also intervenes in the development of normal tissue including hematopoietic cells.[43] The integral membrane protein bcl-2 is produced by proto-oncogene BCL-2 located on chromosome 18, which is found in endoplasmic reticulum membrane, nuclear membrane and outer mitochondrial membrane. Its function is to prevent apoptosis. Overexpression of this protein suppresses apoptosis in damaged cells, contributing to metastases of some tumors.[53] Its expression in fibrous solitary tumors of the pleura has a relationship with the diameter of the tumor: more positive cells equals smaller tumor size. This means that bcl-2 inhibits prolifera-

**Table 2.** Factors associated with prognosis

| Good prognosis | Poor prognosis |
| --- | --- |
| • Peduncle | • No peduncle |
| • Tumors <10 cm | • Tumors >10cm |
| • Encapsulated | • High cellularity |
| • Low pleomorphism | • Accented pleomorphism |
| • Low mitotic activity | • High mitotic activity |
| • Full resection | • Necrosis |
| • No pulmonary invasion | • Hemorrhage |
| • Calcification | • Pleural effusion |
| • Low cellularity | Atypical location |
| • Low proliferative index | • Adjacent structures invasion |
| | • High proliferative index |
| | • p53 overexpression |
| | • CD34− |
| | • Extrathoracic symptoms (hypertrophic osteoarthropathy, hypoglycemia, finger clubbing) |

**Table 3.** Immunohistochemical characteristics

| Marker | % Reactivity |
| --- | --- |
| Vimentin+ | 81-100 |
| CD34+ | 90-95 |
| CD99+ | > 50 |
| bcl-2+ | 50 |
| Kit (CD117)+ | 29-50 |
| SMA+ | 20 |
| Pancytokeratins− | 100 |
| EMA− | |
| S-100 protein− | |
| Desmin− | |
| Calretinin− | |

SMA, smooth muscle actin; EMA, epithelial membrane antigen.

**Table 4.** Literature review

| Author | Briselli, 1981 (rev.)[8] | Briselli, 1981[8] | England, 1989[13] | Weynand, 1997[11] | Suter, 1998[24] | de Perrot, 1999[37] | Galbis Caravajal, 2004[23] | Sánchez-Mora, 2006[6] | Kohler, 2007[38] |
|---|---|---|---|---|---|---|---|---|---|
| Period (years) | 1942-1980 | 1977-1978 | ? | ? | 1967-1997 | 1981-1998 | ? | 1994-2004 | 1993-2006 |
| *n* | 360 | 8 | 223 | 5 | 15 | 10 | 10 | 30 | 27 |
| Country | USA | USA | USA | Belgium | Switzerland | Switzerland | Spain | Spain | Switzerland |
| Gender | 54 % M | 62 % M | 51.6 % M | 80 % M | 60 % F | 80 % M | 80 % M | 67 % F | 52 % F |
| Age (years) | 51 (5-87) | 56 (44-71) | 57 (9-86) | 59-73 | 57 (27-79) | 64 (41-78) | 58.6 (33-76) | 58.39 (18-73) | 62.3 (53-71) |
| Asymptomatic (%) | 36 | 62 | 52 | 80 | 53 | 30 | 80 | 47 | 37 |
| Origin (%) | | | | | | | | | |
| • Visceral pleura | 80 | 50 | 66 | 100 | 67 | 70 | 40 | 83 | 89 |
| • Parenchyma | 0 | 0 | 0 | 0 | 0 | 0 | 10 | 0 | 0 |
| • Parietal pleura | 20 | 50 | 34 | 0 | 33 | 30 | 50 | 17 | 11 |
| Tumor (%) | | | | | | | | | |
| • Solitary | - | 100 | 99 | 100 | 100 | 100 | 100 | 80 | 100 |
| • Multiple | - | 0 | 1 | 0 | 0 | 0 | 0 | 20 | 0 |
| Size (cm) | 6 (1-36) | 11 (1-33) | 5-10 (1-39) | 6.5-19 | ? | 9.8 (7-20) | 8.3 (4-15) | 7.35 (2-21) | 8.2 (2-23) |
| Cellular density (%) | | | | | | | | | |
| • Low | - | 25 | ? | ? | ? | ? | ? | 15 | ? |
| • Moderate | - | 37.5 | ? | ? | ? | ? | ? | 50 | ? |
| • High | - | 37.5 | ? | ? | ? | ? | ? | 35 | ? |
| Atypia (%) | | | | | | | | | |
| • Mild | - | - | ? | ? | ? | 20 | ? | 45 | ? |
| • Moderate | - | - | ? | ? | ? | 10 | ? | 30 | ? |
| • Moderate | - | - | ? | ? | ? | 10 | ? | 15 | ? |
| Necrosis (%) | - | 50 | 33 | ? | ? | 10 | ? | 25 | ? |
| Mitosis > 4/10 HPF (%) | - | 0 | 29 | 20 | ? | 10 | ? | 30 | 37 |

continued

(continuation)

| Author | Briselli, 1981 (rev.)[8] | Briselli, 1981[8] | England, 1989[13] | Weynand, 1997[11] | Suter, 1998[24] | de Perrot, 1999[37] | Galbis Caravajal, 2004[23] | Sánchez-Mora, 2006[6] | Kohler, 2007[38] |
|---|---|---|---|---|---|---|---|---|---|
| IHC(%) | | | | | | | | | |
| • Vimentin+ | - | - | 81 | 100 | 100 | 100 | - | 100 | 100 |
| • CD34+ | - | - | - | - | 84 | 100 | 100 | 85 | 100 |
| • bcl-2+ | - | - | - | - | - | - | - | 65 | - |
| • CD99+ | - | - | - | - | - | - | - | 40 | - |
| • Keratins– | - | - | 100 | 100 | 100 | 100 | - | 100 | 100 |
| • S100 protein– | - | - | 100 | - | - | - | - | 100 | - |
| Follow-up (months) | - | - | 44 (1-372) | ? | 80 (5-324) | 53 (2-168) | 23.9 (6-54) | 66.32 | 54 (6-157) |
| Behavior Malignant behavior (%) | 19 | 25 | 37 | ? | 60 | 20 | 10 | 13 | 37 |
| (hemithorax) | S/P | Left. 75 % | Left. 56 % | Right. 60 % | Right. 47 % | Left. 60 % | ? | - | Left. 59 % |
| Presenting symptoms (%) | | | | | | | | | |
| • Chest pain | 44 | 25 | 55 | 0 | 27 | 50 | 10 | 20 | 26 |
| • Dyspnea | 37 | 12 | 6 | 0 | 13 | 50 | 10 | 20 | 11 |
| • Cough | 46 | 12 | 27 | 0 | 33 | 20 | 0 | 10 | 26 |
| • Hypertrophic osteo-arthropathy | - | - | - | 0 | 6 | 0 | 10 | - | - |
| • Acropachy | - | - | 8 | 20 | 6 | 0 | 0 | - | 7 |

F, female; M, male; R, right; L, left; ?, data not available; HPF, high-power fields; IHC, immunohistochemistry tive cell activity through the interaction of still unknown additional molecules, which indicates that this protein exerts different biological effects according to cell type.[36]

Differential diagnoses include pleural mesothelioma, neurogenic sarcoma, monophasic synovial sarcoma, hemangiopericytoma, fibroblastic sarcoma, and malignant fibrous histiocytoma (determined by immunohistochemistry),[6-7,14] as well as bronchogenic carcinoma,[54] metastatic pulmonary carcinoma,[14] and calcifying fibrous pseudotumor of the pleura.[55] Even though pulmonary sequestration is not included in differential diagnosis, there is frequent misdiagnosis involving this pathology.[56] In our series, one case diagnosed with monophasic synovial sarcoma (through section review) was excluded and required reintervention from remnant tumor at 6 months and at 10 months of metastasis.

Of the patients, 30-60% present symptoms[1,4,9,10,14,57] (Table 5), and this rate reaches 54-67% in benign cases and >75% in malignant cases.[25] Symptoms depend on tumor size and include cough, dyspnea from compressive atelectasis, pleural pain (common if tumor has a parietal pleural origin) and lower limb edema from mediastinum compression.[2,9,26,29] Acropachy and hypertrophic osteoarthropathy secondary to abnormal production of hyaluronic acid from tumor cells[4] (Pierre-Marie-Bamberger syndrome[12]) are found in 20-50% of cases. This is the most common associated paraneoplastic syndrome[25] vs. 6% in patients with mesothelioma[9] and are associated with a malignant phenotype.[10,14] Pleural effusion is uncommon in benign tumors (2%) but may be present in 10-32% of malignant tumors.[2,4,9,13,29,54,58] It is sometimes related with other paraneoplastic syndromes such as secondary reactive hypoglycemia[12,34,59] (Doege-Potter syndrome[12]) from hyperproduction of type 2 insulin-like growth factor (pro-IGF2) precursor in 4-5% of cases (2% in benign cases and 14% in malignant cases[60]). It is more frequent in females and may exceptionally cause hypoglycemic coma[12,61,62] as well as be related to hyponatremia secondary to inadequate secretion of antidiuretic hormone. It may develop with rheumatoid arthritis. All these symptoms disappear after the tumor is removed.[2,4,9,12,25,29] Galactorrhea has also been reported[1,16,25,58] as well as gynecomastia[25] with uncertain etiology. This is an incidental finding in the remaining patients.[2,4,58] Local symptoms are in relation to the large size of the tumor, whereas extrathoracic manifestations are not necessarily related.[33]

In our study we observed symptoms in all patients although dyspnea, coughing and chest pain are the most frequent. Prevalence of all symptoms and sign was higher than documented in larger literature series (Tables 4 and 5), possibly because our series focused on giant tumors.

Fibrous solitary tumors of the pleura cannot be distinguished from other tumors using simple chest x-ray.[1] Contrast CT shows a heterogeneous aspect with a peduncle that usually is attached to visceral pleura,[1,2,9,26,63] also presenting reinforcement from a large vascularization (areas without reinforcement are related with necrosis, hemorrhage or degeneration). Lesions >10 cm in diameter associated with central necrosis and pleural effusion increase probability of malignancy[2] (Table 2). Calcified areas may be found in up to 26% of cases,[25] and in our series these areas were found in 16.7% ($n = 1$) of cases. Multiple pedunculated tumors attached to visceral pleura are rare.[9]

MRI may be useful to evaluate fibrous structure of the tumor.[2,4] Because 79% of lesions extend toward the lower thorax and 28% are adjacent to the diaphragm, both CT and magnetic resonance are useful to evaluate the relationship between tumor and other structures (including mediastinum). These imaging methods can help define the extension of infiltration of adjacent structures when present.[25,57] In this case, MRI shows greater sensitivity.[63]

Ultrasound is not useful for diagnosis[2] but fluorodeoxyglucose positron emission tomography (FDG-PET) has been suggested to identify or discard malignancy prior to surgery.[25,64-66]

Preoperative angiography helps to evaluate vascularization of the tumor,[67-69] observing hypervascularity from aorta-derived arterial contribution and, in some cases, from internal mammary artery.[2] Occasionally the tumor feeds from aberrant blood vessels, phrenic artery branches or bronchial artery branches.[13] Abdominal aorta contribution through renal arteries has been recorded[56] and early venous drainage is typically not observed.[2] In our study we found vascularization from internal mammary artery with a small contribution from branches of bronchial and phrenic arteries.

Correct diagnosis is vital because despite the typically large size of the tumor, the patient has a good prognosis[13,24,33,40] when the tumor is fully resected,[40] observing healing in 45% of cases.[13] However, because there is a potential malignant behavior, all patients (with either benign or malignant tumors[63]) should adhere to a lengthy follow-up program[7,10,11,33] carrying out control CT scan every 6 months during the first 2 years and a yearly control CT scan afterwards[25,37] because most relapses (particularly from sessile malignant tumors ~63%) occur 24 months after resection. Ideally, long-term follow-up should be carried out for 15-20 years.[25]

Giant tumors may generate partial or full atelectasis as well as contralateral mediastinum deviation.[70] This is rarely regarded as a simple case because it generally refers to pleural fibromas.[33] Because of their fibrous and hypocellular characteristics, a transthoracic FNAB offers little value for diagnosis; however, when true-cut[11,33,60] or core biopsy techniques are used, the biopsy increases its diagnostic contribution as demonstrated in our series. Moreover, spe-cialized histopathological techniques such as electron microscopy and immunohistochemical microscopy can increase biopsy specificity.[2,68] Preoperative embolization is useful, especially when dealing with large masses[2,67,69] because this reduces intraoperative bleeding.[56,67]

The treatment of choice is surgical resection[1,4,7,9,14,23,33,57,58,60] of the tumor, its peduncle and site of origin. Other resections are occasionally required: pulmonary, costal wall or en-bloc diaphragm (particularly when tumor originates in the parietal pleura[4]), with 5-year survival rates of 97%.[9] Surgical approaches include thoracotomy, sternotomy and thoracoscopy and are selected according to tumor location and size.[57] Although most tumors are benign, they should be removed with appropriate free surgical margins of 1-2 cm[12,21] because this is the best guarantee for avoiding relapses. When using video-assisted techniques, tumor content effusion should be avoided within the thoracic cavity, particularly when removing the lesion.[4] This is especially useful for treating solitary pulmonary nodules, small fibrous tumors[12,60] (<5 cm[54]) or tumors located within a fissure[60] and mainly when resection is technically feasible and images are clear.[33] This approach has been recommended to evaluate any non-diagnosed tumor through biopsy to determine the best surgical approach and to avoid invasive thoracotomy.[15] However, risk-benefit of this diagnostic procedure should be evaluated because it involves anesthesia and surgery. Overall intraoperative mortality has been estimated at 12% because of hemodynamic changes associated with decompression of mediastinal structures.[67] Postoperative chemo- and radiotherapy have demonstrated no additional benefits[4,7,37,60] even when performing an incomplete resection,[33] which has been associated with low cell content and low mitotic activity of the tumor.[67] However, there is recent reference of cases where chemo- and radiotherapy have been useful.[21,70] A precise preoperative diagnosis using neoadjuvant therapy is difficult even when a biopsy is available.[21,38]

These types of tumors may relapse or transform into a malignancy, usually nonpedunculated, which is larger in size and located in unusual sites. For these cases, the treatment of choice is a second resection, which provides good, long-term patient survival. Incomplete second resection or malignant transformation reduces survival to 24-36 months.[9] On the other hand, repeated resections to treat relapses increase technical difficulty.[40] Relapses have been reported with a prevalence of 1.4-23%[6,17,30,36,54] distant metastases with 0-19% and tumor-related mortality of 0-27%.[6,13] Relapse and metastases generally occur 5 years after surgery[7] although latency periods up to 31 years have been reported.[13,10,11] Malignant transformation is unusual;[17] however, malignant fibrous histiocytoma transformations have been reported.[71] Local relapse after full resection is not uncommon in malignant tumors but is exceptionally rare in benign lesions. This may be a consequence of incomplete resection, unidentified tumor during surgery or growth of a nonrelated second tumor.[38] However a "contact" mechanism has been suggested between these.[72] De Perrot et al.[21] esta-blished a classification in 2002 to stratify relapse risk based on histological and morphological markers and a management algorithm (Figure 3). They proposed a classification for these types of tumors: stage 0—pedunculated tumor without signs of malignancy; stage I—sessile (inverted) tumor without malignancy signs; stage II—pedunculated tumor with signs of histological malignancy; stage III—sessile (inverted) tumor with histological malignancy signs; stage IV—multiple synchronous metastatic tumors.

**Table 5.** Associated symptomatology

| Symptom | Okike, 1978[62] | England, 1989[13] | Suter, 1998[24] | de Perrot, 1999[37] | Rena, 2001[63] | Magdeleina, 2002[54] | Sung, 2005[5] | Carretta, 2006[41] | Kohler, 2007[38] | Santolaya, 2007[28] |
|---|---|---|---|---|---|---|---|---|---|---|
| n | 52 | 223 | 15 | 10 | 21 | 60 | 63 | 18 | 27 | 41 |
| Asymptomatic (%) | 54 | 52 | 53 | 30 | 57 | 48 | 43 | 56 | 37 | 24 |
| Chronic cough (%) | 33 | 27 | 33 | 20 | 14 | 32 | 8 | 17 | 26 | 29 |
| Pleural pain (%) | 23 | 55 | 27 | 50 | 28 | 48 | 25 | 6 | 26 | 24 |
| Dyspnea (%) | 19 | 6 | 13 | 50 | 19 | 13 | 17 | 17 | 11 | 19 |
| Fever (%) | 17 | 3 | - | - | - | - | 2 | 11 | - | - |
| Hypertrophic pulmonary osteo-arthropathy with or without acropachy (%) | 19 | 8 | 13 | - | 14 | 13 | 2 | 6 | 7 | - |
| Pleural effusion (%) | 6 | 17 | - | - | 19 | 23 | - | - | 4 | 17 |
| Weight loss (%) | 6 | 10 | - | 10 | 14 | - | 3 | 6 | 7 | - |
| Hemoptysis (%) | 2 | 3 | 7 | 10 | - | - | - | - | - | - |
| Pneumonitis (%) | 2 | - | - | 10 | - | - | - | - | 7 | - |

In conclusion, giant solitary fibrous tumors of the pleura deserve special consideration. Because of their size they may trigger several symptoms in patients and surgical removal is technically more difficult. It is advisable to carry out preoperative embolization in order to reduce intraoperative bleeding. Scarcity of literature in regard to these large tumors was an obstacle in carrying out an appropriate comparison. Most often they are mentioned in series that include lesions of different sizes and are case reports. Further studies with additional patients are needed as well as a longer follow-up period in order to obtain data with higher statistical power.

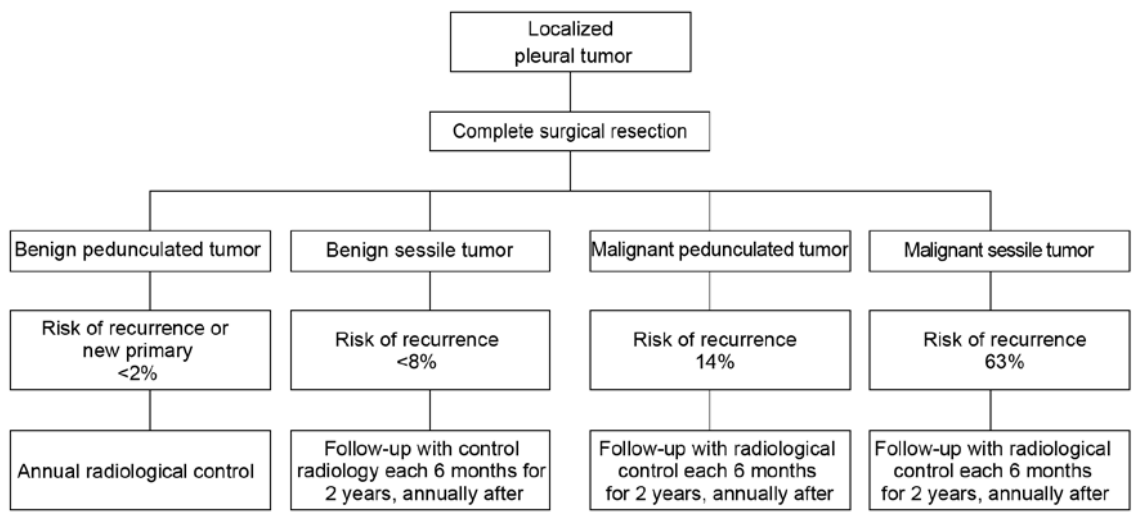

**Figure 3.** Management algorithm proposed by de Perrot et al. Modified from Reference 21.